# Quantum Oscillations of the Critical Current of Asymmetric Aluminum Loops in Magnetic Field

V.L. Gurtovoi, S.V Dubonos, A.V. Nikulov, N.N. Osipov, and V.A. Tulin

*Institute of Microelectronics Technology and High Purity Materials, Russia Academy of Sciences, 142432 Chernogolovka, Moscow Region, Russia. E-mail: nikulov@ipmt-hpm.ac.ru*

**Abstract.** The periodical dependencies in magnetic field of the asymmetry of the current-voltage curves of asymmetric aluminum loop are investigated experimentally at different temperatures below the transition into the superconducting state $T < T_c$. The obtained periodical dependency of the critical current on magnetic field allows to explain the quantum oscillations of the dc voltage as a consequence of the rectification of the external ac current and to calculate the persistent current at different values of magnetic flux inside the loop and temperatures.

## INTRODUCTION

The persistent current $I_p = sj_p = s2en_sv_s$ should flow along circumference of a superconductor loop l with thin section $s < \lambda_L^2$ at $\Phi \neq n\Phi_0$ since the state with zero velocity of superconducting pairs $v_s = 0$ is forbidden when the magnetic flux $\Phi$ inside l is not divisible by the flux quantum $\Phi_0 = h/2e$ [1]. Its equilibrium value and sign $<I_p> = s2en_s<v_s> \propto <n> - \Phi/\Phi_0$ vary periodically with $\Phi$ [1]. The Little-Parks resistance oscillations $R_l(\Phi/\Phi_0)$ [2] observed in the loop [3] is experimental evidence of $I_p \neq 0$ at non-zero resistance $R_l > 0$. According to an analogy with the conventional circular current a dc potential difference $V(\Phi/\Phi_0) \propto I_p(\Phi/\Phi_0) \propto <n>-\Phi/\Phi_0$ may be expected to be observed on segments of asymmetric superconductor loop at $R_l > 0$. Such quantum oscillations of the dc voltage were observed on segment of asymmetric aluminum loops [4,5] and much before on a double Josephson point contact [6]. The dc voltage $V(\Phi/\Phi_0)$ observed near the critical temperature $T_c$ [4,6] can be induced by switching of the loop between superconducting states with different connectivity [7,8]. The quantum oscillations $V(\Phi/\Phi_0)$ induced at lower temperatures by an external ac current [5] may by interpreted as a result of the rectification because of asymmetry of the current-voltage curves sign and value of which are periodical function of $\Phi$. The results of measurements of this periodical change of the asymmetry of the current-voltage curves with value of magnetic flux $\Phi$ inside asymmetric aluminum loop at different temperatures are presented in this work.

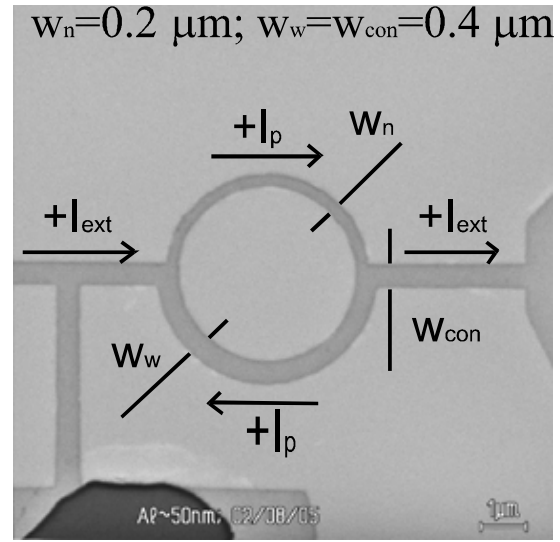

**FIGURE 1.** SEM image of a typical asymmetric aluminium rings.

## EXPERIMENTAL DETAILS AND RESULTS

Microstructures consisting of asymmetric Al rings with semi-ring width $w_n = 200$ nm and $w_w = 400$ nm for the narrow and wide parts, respectively, see Fig.1, were investigated. 4 μm diameter single asymmetric superconductor ring (ASR) and 20 ASR structures, see Fig.1, were fabricated by e-beam lithography and lift-off process of film d = 45-50 nm in thickness, thermally evaporated on oxidized Si substrates. For these structures, the sheet resistance was 0.23 Ω/□ at 4.2 K, the resistance ratio R(300K)/R(4.2K)=2.7, and the critical temperature was $T_c$ = 1.24-1.27 K.

## Current -Voltage Curves

The structure as a whole jumps into the resistive state R > 0 (at a low temperature T < 0.985$T_c$) when the current density exceeds the critical value $j_c$ in any of its segment and the irreversibility of the current-voltage curves is observed at T < 0.99$T_c$. The value of the external current $I_{ext}$ corresponding to this jump to the state with R > 0 is measured as the critical current $|I_{ext}|_c$ of the structure.

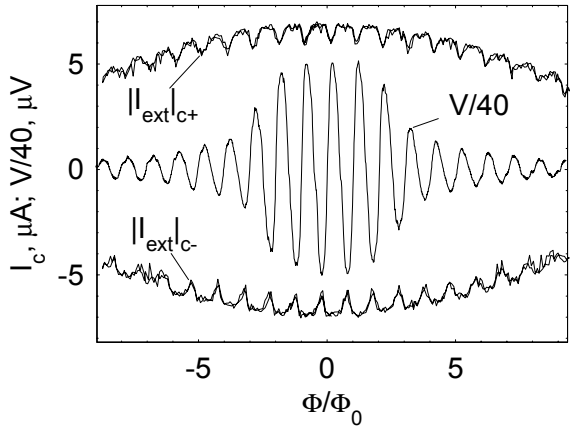

**FIGURE 2.** Quantum oscillations of the critical current $|I_{ext}|_c(\Phi/\Phi_0)$ of system of 18 asymmetric Al loops with $w_{con}$ = 400 nm < $w_n + w_w$ measured in opposite directions $|I_{ext}|_{c+}$, $|I_{ext}|_{c-}$ at T ≈ 0.978$T_c$. The quantum oscillations of the dc voltage $V(\Phi/\Phi_0)$ induced by the external ac current with the frequency f = 40 kHz and the amplitude 7 μA are shown also.

## Periodical Dependence of $|I_{ext}|_c(\Phi/\Phi_0)$

Our measurements have revealed the periodical magnetic dependencies $|I_{ext}|_c(\Phi/\Phi_0)$ of both single rings and systems of rings with both $w_{con} < w_n + w_w$, see Fig.2, and $w_{con} \geq w_n + w_w$. These periodical dependencies may be explained as a consequence of superposition of the external $I_{ext}$ and persistent $I_p(\Phi/\Phi_0)$ currents. The current density in the narrow $j_n$ and wide $j_w$ semi-rings is determined by both the $I_{ext}$ and $I_p$ currents: $j_n = I_{ext}/d(w_n + w_w) \pm I_p/dw_n$ and $j_w = I_{ext}/d(w_n + w_w) \pm I_p/dw_w$ [5]. The summation "+" takes place when the direct $I_{ext}$ and circular $I_p$ currents have the same direction in the semi-rings, see Fig.1. The current density mounts the critical value $j_c$ first of all in the narrow semi-rings at $|I_{ext}|_c = d(w_n + w_w)(j_c - I_p/w_n)$, in the wide ones at $|I_{ext}|_c = d(w_n + w_w)(j_c - I_p/w_w)$ or in the stripes connecting the rings at $|I_{ext}|_c = dw_{con} j_c$ depending on the $I_{ext}$ and $I_p$ directions and the $j_p/j_c$ and $w_{con}/(w_n + w_w)$ values.

Our results of the measurement of $|I_{ext}|_c(\Phi/\Phi_0)$ are evidence of the periodical dependence not only value but also sign of $I_p(\Phi/\Phi_0)$ since the $|I_{ext}|_c(\Phi/\Phi_0)$ value depends on the $I_{ext}$ direction, $|I_{ext}|_{c+} \neq |I_{ext}|_{c-}$, i.e. the current-voltage curves are asymmetric, at some $\Phi/\Phi_0$ values, see Fig.2. For example $|I_{ext}|_{c+}$ (at the right $I_{ext}$ direction) has a minimum value at $\Phi/\Phi_0 \approx 0.2 \pm n$ whereas $|I_{ext}|_{c-}$ (at the left $I_{ext}$ direction) the minimum is observed at $\Phi/\Phi_0 \approx 0.8 \pm n$, see Fig.2. This means that the persistent current has the clockwise direction at $\Phi/\Phi_0 \approx 0.2 \pm n$ and the counter-clockwise one at $\Phi/\Phi_0 \approx 0.8 \pm n$ since the minimum of $|I_{ext}|_c$ is observed when the $I_{ext}$ and $I_p$ have the same direction in the narrow semi-rings, see Fig.1. The experimental dependencies $|I_{ext}|_{c+}(\Phi/\Phi_0)$, $|I_{ext}|_{c-}(\Phi/\Phi_0)$ obtained in our work allow to explain the quantum oscillations of the dc voltage $V(\Phi/\Phi_0)$ as consequence of the rectification and to calculate the $I_p(\Phi/\Phi_0;T)$ dependencies at T < $T_c$.

## ACKNOWLEDGMENTS

This work has been supported by a grant of the Program "Low-Dimensional Quantum Structures", the Russian Foundation of Basic Research (Grant 04-02-17068) and a grant of the program "Technology Basis of New Computing Methods".

## REFERENCES

1. M. Tinkham, *Introduction to Superconductivity,* New York: McGraw-Hill Book Company, 1975.
2. W.A. Little and R.D. Parks, *Phys. Rev. Lett.* **9**, 9 (1962).
3. H. Vloeberghs et al., *Phys. Rev. Lett.* **69**,1268 (1992).
4. S.V. Dubonos, V.I. Kuznetsov, and A.V. Nikulov, in *Proceedings of 10th International Symposium "NANOSTRUCTURES: Physics and Technology"*, St Petersburg: Ioffe Institute, 2002, pp. 350-354; e-print arXiv: cond-mat/0305337.
5. S.V. Dubonos et al., *Pisma Zh. Eksp. Teor. Fiz.* **77,** 439 (2003) *(JETP Lett.* **77,** 371 (2003) ).
6. A. Th. A. M. de Waele et al., *Physica* **37,** 114 (1967).
7. A.V. Nikulov and I.N. Zhilyaev, *J. Low Temp. Phys.* **112**, 227-236 (1998).
8. A.V. Nikulov, *Phys. Rev. B* **64**, 012505 (2001).